\documentclass[pra,twocolumn,superscriptaddress,showpacs,preprintnumbers,amsmath,amssymb]{revtex4}
\usepackage{bm}
\usepackage{bbm}
\usepackage{amssymb}
\usepackage{amsfonts}
\usepackage{epsfig,graphicx}
\usepackage{amstext}
\usepackage{amsmath}
\usepackage{graphicx}
\usepackage{times}
\usepackage{txfonts}
\usepackage{dcolumn}

\begin{document}

\title{Quantum coherence of multiqubit states in correlated noisy channels}

\author{Ming-Liang Hu}
\email{mingliang0301@163.com}
\affiliation{School of Science, Xi'an University of Posts and Telecommunications, Xi'an 710121, China}
\affiliation{Institute of Physics, Chinese Academy of Sciences, Beijing 100190, China}
\author{Heng Fan}
\email{hfan@iphy.ac.cn}
\affiliation{Institute of Physics, Chinese Academy of Sciences, Beijing 100190, China}
\affiliation{CAS Center for Excellence in Topological Quantum Computation, University of Chinese Academy of Sciences, Beijing 100190, China}
\affiliation{Songshan Lake Material Laboratory, Dongguan 523000, China}

\begin{abstract}
The long-time maintenance of quantum coherence is crucial for its
practical applications. We explore decoherence process of a
multiqubit system passing through a correlated channel (phase flip,
bit flip, bit-phase flip, and depolarizing). The results show that
the decay of coherence was evidently delayed when the consecutive
actions of the channel on the sequence of qubits has some classical
correlations. In particular, the relative entropy of coherence for a
system with large number of qubits is more robust than that with
small number of qubits. We also provide an explanation for the
delayed decoherence by exploring the interplay between the change of
the unlocalized quantum coherence and the total correlation gain of
the multiqubit system.
\end{abstract}

\pacs{03.67.Mn, 03.65.Ta, 03.65.Yz}

\maketitle

\section{Introduction} \label{sec:1}
Quantum coherence originates from the superposition principle of
states. It captures an essential nature of quantum systems which is
different from the classical coherence phenomena \cite{Ficek}. Due
to this nature, quantum coherence attracted wide interests of
researchers \cite{Plenio,Hu}. One of such interest is to seek
general conditions under which this genuine quantum property can be
sustained for a system subject to typical sources of noise. The
reasons for such an interest is that quantum coherence can be
exploited as a resource for many quantum information and
communication tasks \cite{Nielsen}. Recent studies also provide
evidence that the long-lived quantum coherence is crucial for
defeating the classical limits of measurement accuracy in quantum
metrology \cite{metro1,roc1,subc} and for enhancing certain
biological processes \cite{biology}. But in practice, the
unavoidable interaction of a principal system with its surroundings
results in decoherence in most situations. As a consequence, the
advantage of quantum coherence to empower the performance of quantum
tasks will disappear in a very short time interval, and this remains
one of the major obstacles for its wide applications.

Due to the above reasons, making clear decoherence mechanism of a
principal system in the presence of typical noises becomes one of
the major goals for accomplishing new quantum technologies. In 2014,
Baumgratz \textit{et al.} \cite{coher} introduced a rigorous
framework for quantifying coherence and defined the $l_1$ norm of
coherence and relative entropy of coherence. Subsequently, many
other faithful coherence measures such as the entanglement-based
coherence measures \cite{coher-ent}, the robustness of coherence
\cite{roc1,roc2}, and the convex roof measures of coherence
\cite{convex1,convex2, convex3}, have also been introduced. All
these paved the way for a quantitative analysis of the decoherence
process of a system coupled to its surrounding environments or under
the action of specific quantum channels.

Based on the measures of coherence, much endeavors have been devoted
to clarifying decoherence mechanism of an open quantum system. The
attempts to achieve this goal were carried out mainly along two
different but closely related directions. The first one is aimed at
explaining role of active quantum operations on coherence of a state
\cite{lqo1,lqo2,lqo3,lqo4, lqo5}, the cohering and decohering power
of specific quantum channels \cite{power1,power2,power3,power4,
power5, power6}, the coherence-breaking channels \cite{break}, and
the non-coherence-generating channels \cite{ncgc}. There are also
some works concentrating on the steered coherence at one party of a
bipartite state by local operations on its another party
\cite{steer,naqc1,naqc2,naqc3}. The second direction is focused on
exploring coherent properties of a principal system coupled to
environment. Along this direction, the frozen phenomenon of
coherence \cite{fro1,fro2,fro3,fro4} and the factorization relation
for coherence evolution \cite{fact} have been observed. Several
theoretical studies focused on suppressing the undesirable decay of
coherence \cite{dyna1,dyna2,dyna3} or generating steady-state
coherence \cite{dyna4,dyna5} have also been performed.

While the above works studied only the case for which the channel
acts identically and independently on the sequence of qubits passing
through it, the effects of correlations on consecutive applications
of the channel on coherence of these qubits is usually neglected.
But this is indeed a very realistic problem which should be taken
into account in experiments. For example, when $N$ qubits traverse a
channel, apart from the very special case for which the time
interval between its successive applications on the qubits is
infinitely small, the channel will has partial memory about its past
history and the uncorrelated channel model will not be applicable
\cite{cc0}. Especially, the correlated applications of the channel
and the temporal correlations in the evolution process of each qubit
may together affect decoherence process of a multiqubit system.

In this paper, we make a step toward the above problem. We consider
a model with classical correlations between consecutive applications
of a channel \cite{cc1}, and examine how the coherence of a
multiqubit state changes when one varies correlation strength of the
channel. This can help to confirm potential role of the classical
correlation of a correlated channel on suppressing decoherence. We
also explore interplay between the extra correlations created by the
channel and change of unlocalized quantum coherence for the
principal system, aimed at providing an explanation for the delayed
decoherence. As any physical process (time evolution, quantum
operation, etc.) can be represented as a channel transforming the
input states into output ones, the decohering channel model
considered here is rather general, and the obtained results are
hoped to shed some light on understanding decoherence process of a
multipartite system immersed in real environments.

\section{Preliminaries} \label{sec:2}
We start by collecting some preliminaries that we employ in this
paper. For the purpose of providing a comparative study, we use two
different coherence measures. The first one is the $l_1$ norm of
coherence defined as \cite{coher}
\begin{equation}\label{eq2-1}
 C_{l_1}(\rho)= \min_{\delta\in \mathcal {I}}\|\rho-\delta\|_{l_1}
              = \sum_{i\neq j}|\langle i|\rho|j\rangle|,
\end{equation}
where $\mathcal{I}$ denotes the set of incoherent states which are
diagonal in the prefixed reference basis $\{|i\rangle\}$.

The second measure we adopt is the relative entropy of coherence
given by \cite{coher}
\begin{equation}\label{eq2-2}
 C_r(\rho)= \min_{\delta\in \mathcal{I}}S(\rho\|\delta)
          = S(\Delta(\rho))-S(\rho),
\end{equation}
with $\Delta(\rho)=\sum_i \langle i|\rho|i\rangle |i\rangle\langle
i|$ denoting the full dephasing of $\rho$ in the basis
$\{|i\rangle\}$, and $S(\cdot)$ is the von Neumann entropy.

Next, we recall the model characterizing correlated channel that we
intend to adopt \cite{cc1}. It provides a proper mathematical tool
apt to describe classical correlations between consecutive
applications of the channel on a sequence of quantum systems
\cite{cc0}. For a single-qubit state $\rho_0$, the common action of
a channel $\mathcal {E}$ on it can be described by a random rotation
of it. That is, $\rho= \sum_{i=0}^3 E_i\rho_0 E_i^\dag$, where the
Kraus operators $E_i=\sqrt{p_i}\sigma_i$, with $\sigma_0=\openone$
being the identity operator, $\sigma_{1,2,3}$ being the Pauli
operators, and $p_i$ constitute a probability distribution.
Similarly, if $\rho_0$ is a $N$-qubit state and $\mathcal {E}$ acts
identically and independently on each of the qubit, the output state
will be given by
\begin{equation}\label{eq2-3}
 \rho= \sum_{i_1 i_2 \cdots i_N} E_{i_1 i_2 \cdots i_N}\rho_0 E_{i_1 i_2 \cdots i_N}^\dag,
\end{equation}
where $E_{i_1 i_2 \cdots i_N}=\sqrt{p_{i_1 i_2 \cdots i_N}}
\sigma_{i_1}\otimes\sigma_{i_2}\otimes\cdots \otimes\sigma_{i_N}$,
and $p_{i_1 i_2 \cdots i_N}=p_{i_1} p_{i_2} \cdots p_{i_N}$
represents the joint probability that a random sequence of Pauli
rotations around the $i_1\cdots i_N$ axes is applied to the sequence
of $N$ qubits traversing the channel.

Apart from the above case, one may encounters the case for which
there are classical correlations on consecutive applications of the
channel, and this may modify the way it acts on an input state. To
be explicit, the action of the channel on a qubit may affects
probability of its action on the subsequent qubits \cite{cc0}.
Macchiavello and Palma \cite{cc1} proposed such a model, for which
the joint probability is given by
\begin{equation}\label{eq2-4}
 p_{i_1 i_2 \cdots i_N}=p_{i_1}p_{i_2|i_1}\cdots p_{i_N|i_{N-1}},
\end{equation}
with $p_{i_n|i_{n-1}}=(1-\mu)p_{i_n}+\mu\delta_{i_ni_{n-1}}$. Here,
$\mu\in[0,1]$ is a parameter which can be interpreted as the
probability that the same Pauli transformation is applied to the
qubits $i_{n-1}$ and $i_n$. For $\mu=0$, this model describes
independent applications of the channel, while for $\mu=1$, the
applications of the channel turn to be fully correlated \cite{cc1}.

In this paper, we focus our attention on the bit flip, bit-phase
flip, phase flip, and depolarizing channels, which belong to the
class of Pauli channels described by Eq. \eqref{eq2-3}. The nonzero
elements of $\{p_i\}$ for the former three channels can be
represented by $p_0=1-p$ and $p_i=p$ (with $i=1$, 2, and 3,
respectively), while those for the last one are represented by
$p_0=1-p$ and $p_{1,2,3}=p/3$ \cite{cc0,cc1,cc2}.

\section{Quantum coherence in correlated quantum channels} \label{sec:3}
We take the maximally coherent state $|\Psi_N\rangle=
|\Psi_1\rangle^{\otimes N}$ as the input of the channel, where
$|\Psi_1\rangle=(|0\rangle+|1\rangle)/\sqrt{2}$. Due to possible
memory effects arising dynamically during the evolution of each
qubit and the memory effects induced by correlated actions of the
channel \cite{cc2}, rich phenomena of decoherence process may be
observed. For the purpose of comparing decoherence rate of the
system with different number of qubits, we use in the following the
normalized version of coherence
\begin{equation}\label{eq3-1}
 \tilde{C}_{\alpha}(\rho)= \frac{C_{\alpha}(\rho)}{C_{\alpha}(|\Psi_N\rangle)},
\end{equation}
with $\alpha=\{l_1,re\}$, $C_{l_1}(|\Psi_N\rangle)=2^N-1$, and
$C_{re}(|\Psi_N\rangle)=N$.

\subsection{Phase flip channel}
First, we consider the $l_1$ norm of coherence. For the uncorrelated
and fully correlated phase flip channels, we have
\begin{equation}\label{eq3a-1}
 \tilde{C}_{l_1}(\rho)=\left\{
  \begin{aligned}
   & \frac{\sum_{n=1}^N \alpha_n^{(N)} |1-2p|^n}{2^N-1}  && \mathrm{for}~ \mu=0, \\
   & \frac{2^{N-1}(1+|1-2p|)-1}{2^N-1}               && \mathrm{for}~ \mu=1,
  \end{aligned} \right.
\end{equation}
where the coefficients $\alpha_1^{(N)}=N$, $\alpha_N^{(N)}=1$, and
$\alpha_n^{(N)}=\alpha_{n-1}^{(N-1)}+\alpha_{n}^{(N-1)}$ for
$n\in[2,N-1]$.

\begin{figure}
\centering
\resizebox{0.47 \textwidth}{!}{%
\includegraphics{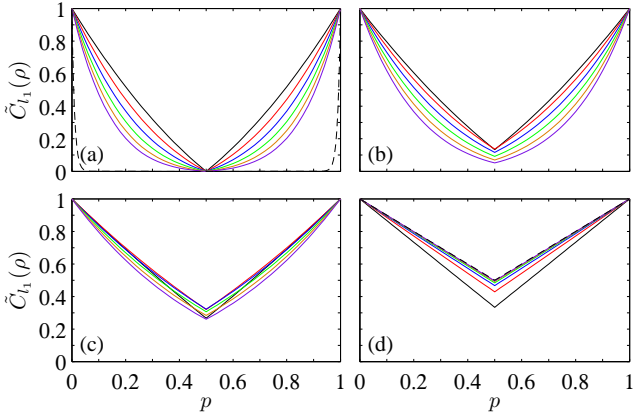}}
\caption{The $p$ dependence of $\tilde{C}_{l_1}(\rho)$ for the input
state $|\Psi_N\rangle$ and the correlated phase flip channel with
(a) $\mu=0$, (b) $\mu=0.4$, (c) $\mu=0.8$, and (d) $\mu=1$. The
solid black, red, blue, green, orange, and purple lines correspond
to $N=2$, 3, 4, 5, 6, and 7, respectively. The dashed lines in (a)
and (d) correspond to $N=100$ and $N\rightarrow \infty$,
respectively.} \label{fig:1}
\end{figure}

Based on Eq. \eqref{eq3a-1}, one can see that $\tilde{C}_{l_1}
(\rho)$ is symmetric with respect to $p=0.5$. For both the cases
$p=0$ and 1, $\tilde{C}_{l_1}(\rho)=1$, while for $p=0.5$,
$\tilde{C}_{l_1}(\rho)$ takes its minimum which is zero for $\mu=0$
and finite for $\mu=1$. In the region of $p\leqslant 0.5$,
$\tilde{C}_{l_1}(\rho)$ decreases with the increasing $p$. Moreover,
for the uncorrelated channel ($\mu=0$) with $p\neq \{0,1\}$,
$\tilde{C}_{l_1}(\rho)$ decreases with the increasing number $N$ of
qubits  and vanishes when $N\rightarrow \infty$. See, e.g., the
dashed line displayed in Fig. \ref{fig:1}(a). For the fully
correlated channel ($\mu=1$), however, $\tilde{C}_{l_1}(\rho)$ turns
to be increased from $|1-2p|$ to $(1+|1-2p|)/2$ when $N$ increases
from 1 to $\infty$. In particular, from Fig. \ref{fig:1}(d) one can
observe that the lines of $\tilde{C}_{l_1}(\rho)$ with $N=7$ and
$N\rightarrow \infty$ are already nearly overlapped. This indicates
that the fully correlated actions of the phase flip channel can
significantly enhance the $l_1$ norm of coherence in the whole time
evolution process.

For the intermediate value $\mu=0.5$ and $p\leqslant 0.5$ [for
$p>0.5$, $\tilde{C}_{l_1}(\rho)$ can be obtained by substituting $p$
with $1-p$], we obtain
\begin{equation}\label{eq3a-2}
 \tilde{C}_{l_1}(\rho)=\frac{\sum_{n=1}^{N+1} (-1)^{n-1} \beta_n^{(N)} p^{n-1}}{2^N-1},
\end{equation}
with $\beta_1^{(N)}=2^N-1$, $\beta_2^{(N)}= 2^{N-1} (N+1)$, and
$\beta_n^{(N)}=2\beta_{n}^{(N-1)} +\beta_{n-1}^{(N-1)}$ for
$n\in[3,N+1]$. For any fixed $N$, $\tilde{C}_{l_1}(\rho)$ decreases
monotonically when $p$ increases from 0 to 0.5, and apart from
$p=\{0,1\}$, $\tilde{C}_{l_1}(\rho)\rightarrow 0$ when the qubit
number $N\rightarrow \infty$. For general values of $\mu$, we
performed numerical calculation. The results show that the
dependence of $\tilde{C}_{l_1}(\rho)$ on $p$ is qualitatively the
same to that of $\mu=0.5$. For fixed $p$, $\tilde{C}_{l_1}(\rho)$
decreases with the increasing $N$ for relative small $\mu$, see Fig.
\ref{fig:1}(b). But as can be seen from Fig. \ref{fig:1}(c),
$\tilde{C}_{l_1}(\rho)$ may does not always behave as a monotonic
function of $N$ for relative large $\mu$.

From the above analysis one can see that correlated actions of the
phase flip channel is beneficial for long-time preservation of the
$l_1$ norm of coherence. We have performed numerical calculations
with different $p$, and found that for the input state
$|\Psi_N\rangle$, $\tilde{C}_{l_1}(\rho)$ always behaves as a
monotonic increasing function of $\mu$. For $p=0.5$, we managed to
obtain its analytical solution as
\begin{equation}\label{eq3a-3}
  \tilde{C}_{l_1}(\rho)=\frac{1}{2^N-1}\sum_{n=1}^{N-1} \eta_n^{(N)}\mu^n,
\end{equation}
where $\eta_1^{(N)}=N-1$, $\eta_{N-1}^{(N)}=1$, and $\eta_n^{(N)}=
\eta_{n-1}^{(N-1)}+\eta_{n}^{(N-1)}$ for $ n\in [2,N-2]$. By using
this equation, one can confirm again that $\tilde{C}_{l_1}(\rho)$ is
a monotonic increasing function of $\mu$. That is, the $l_1$ norm of
coherence can be enhanced by introducing classical correlation to
this channel. But apart from the special case $\mu=1$ for which
$\tilde{C}_{l_1}(\rho)$ always takes a finite value, the enhancement
will becomes smaller and smaller with the increasing number $N$ of
qubits, and $\tilde{C}_{l_1}(\rho) \rightarrow 0$ when $N\rightarrow
\infty$. Further numerical calculation shows that the similar
phenomena also happen for general values of $p\neq \{0,1\}$.

\begin{figure}
\centering
\resizebox{0.47 \textwidth}{!}{%
\includegraphics{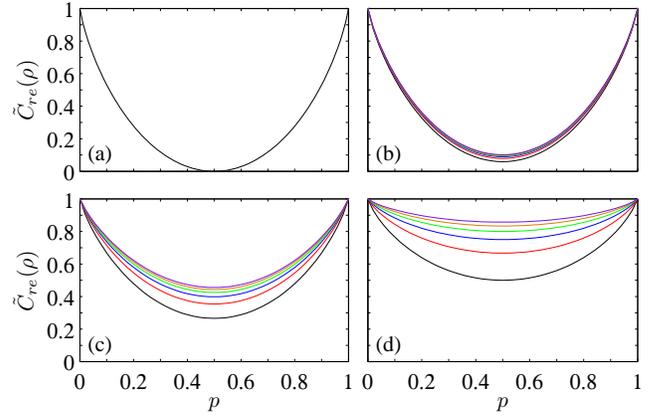}}
\caption{The $p$ dependence of $\tilde{C}_{re}(\rho)$ for the input
state $|\Psi_N\rangle$ and the correlated phase flip channel with
(a) $\mu=0$, (b) $\mu=0.4$, (c) $\mu=0.8$, and (d) $\mu=1$. The
solid black, red, blue, green, orange, and purple lines correspond
to $N=2$, 3, 4, 5, 6, and 7, respectively.} \label{fig:2}
\end{figure}

Next, we turn to consider the relative entropy of coherence. For the
uncorrelated and fully correlated phase flip channels, analytical
expressions of $\tilde{C}_{re}(\rho)$ can be obtained as
\begin{equation}\label{eq3a-4}
 \tilde{C}_{re}(\rho)=\left\{
  \begin{aligned}
   & 1-H_2(p)             && \mathrm{for}~ \mu=0, \\
   & 1-\frac{1}{N}H_2(p)  && \mathrm{for}~ \mu=1,
  \end{aligned} \right.
\end{equation}
with $H_2(\cdot)$ being the binary Shannon entropy.
$\tilde{C}_{re}(\rho)$ decreases monotonically when $p$ increases
from 0 to 0.5, see Fig. \ref{fig:2}(a) and (d). For $\mu=0$,
$\tilde{C}_{re}(\rho)$ is independent of the number $N$ of qubits,
while for $\mu=1$, it increases with the increase of $N$ and
approaches asymptotically to its maximum 1 when $N \rightarrow
\infty$. This is a surprising phenomenon. It indicates that the
relative entropy of coherence for a system with large number of
qubits has an intrinsic rigidity against the fully correlated phase
flip channel. This phenomenon is also in big contrast to that of the
$l_1$ norm of coherence, as apart from the trivial cases $p=0$ and
1, the latter can only approaches to a finite value other than the
maximum 1 in the limit of $N \rightarrow \infty$.

For $0<\mu<1$, a further analysis shows that $\tilde{C}_{re}(\rho)$
exhibits a linear dependence on $1/N$. To be explicit, we have
\begin{equation}\label{eq3a-5}
 \tilde{C}_{re}(\rho)= 1-H_2(p)+ k H_2(p)\left(1-\frac{1}{N}\right),
\end{equation}
and the factor $k$ can be evaluated analytically as
\begin{equation}\label{eq3a-6}
 k=\frac{\sum_n \epsilon_n \log_2 \epsilon_n +2H_2(p)}{H_2(p)}.
\end{equation}
where $\epsilon_{1,2}=p(1-p)(1-\mu)$, $\epsilon_3= p(p+\mu-p\mu)$,
and $\epsilon_4=(1-p)(1-p+p\mu)$ are the eigenvalues of
$\mathcal{E}(|\Psi_2\rangle)$.

From Eq. \eqref{eq3a-5} one can verify directly that the $p$
dependence of $\tilde{C}_{re}(\rho)$ is qualitatively the same to
that of $\mu=1$. That is, it decreases monotonically when $p$
increases from 0 to 0.5, see the exemplified plots displayed in Fig.
\ref{fig:2}(b) and (c). Moreover, $\tilde{C}_{re}(\rho)$ increases
monotonically with the increase of $\mu$, i.e., the decay of the
relative entropy of coherence can be suppressed evidently by the
classical correlations arising from the consecutive applications of
the phase flip channel.

\begin{figure}
\centering
\resizebox{0.46 \textwidth}{!}{%
\includegraphics{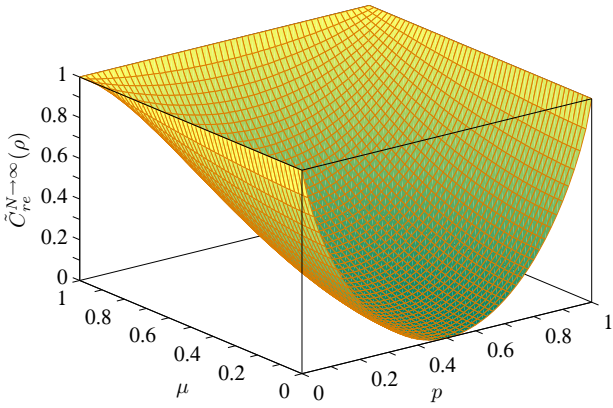}}
\caption{The $p$ and $\mu$ dependence of $C_{re}^{N\rightarrow
\infty}(\rho)$ for the input state $|\Psi_N\rangle$ and the
correlated phase flip channel.} \label{fig:3}
\end{figure}

From Eq. \eqref{eq3a-5} one can also see that $\tilde{C}_{re}(\rho)$
always increases with the increasing qubit number $N$. This is an
important feature different from that exhibited by the $l_1$ norm of
coherence. It shows again that the relative entropy of coherence
with large number of qubits is more robust than that with small
number of qubits, and reflects from one aspect the difference
between these two measures of coherence \cite{ordering}. When
$N\rightarrow \infty$, $\tilde{C}_{re}(\rho)$ approaches to its
asymptotic value $C_{re}^{N\rightarrow \infty}(\rho)$. This
asymptotic value increases from $1-H_2(p)$ to its maximum 1 when
$\mu$ increases from 0 to 1 (see Fig. \ref{fig:3}). More
importantly, $C_{re}^{N\rightarrow \infty}(\rho)$ is finite apart
from the very special point $(\mu,p)=(0,0.5)$. This further confirms
our finding that the relative entropy of coherence for a system with
large number of qubits has an intrinsic rigidity against the
correlated phase flip channel.

Before ending this section, we give a short comment of the facts
that when $N\rightarrow \infty$, $\tilde{C}_{l_1} (\rho) \rightarrow
(1+|1-2p|)/2$ and $\tilde{C}_{re}(\rho) \rightarrow 1$ for $\mu=1$,
while $\tilde{C}_{l_1}(\rho)\rightarrow 0$ and $\tilde{C}_{re}
(\rho)\rightarrow 1+(k-1)H_2(p)$ for $0<\mu<1$. At first glance, it
seems to be contradictory, as any coherence measure vanishes for
incoherent states and takes its maximum for maximally coherent
states. But in fact there is no such a contradiction. This is
because $\tilde{C}_{re}(\rho) \rightarrow 1$ indicates
$C_{re}(\rho)$ is infinitely close to its maximum $N$, while the
output state $\rho$ is still not maximally coherent. Similarly,
$\tilde{C}_{l_1}(\rho) \rightarrow 0$ indicates $C_{l_1}(\rho)$ is
far less than its maximum $2^N-1$, while the output state $\rho$
still maintains partial coherence.

\subsection{Bit flip, bit-phase flip, and depolarizing channels}
For these types of noisy channels, the considered coherence can be
obtained directly based on the results for the correlated phase flip
channel. We list them as follows:

(\romannumeral+1) For the correlated bit flip channel, it is direct
to obtain that $|\Psi_N\rangle$ is an eigenstate of
$\sigma_1^{\otimes N}$ and, therefore, the $N$ qubits will pass
undisturbed through this channel. As a consequence, the quantum
coherence will frozen forever.

(\romannumeral+2) For the correlated bit-phase flip channel, as
$\sigma_2= i\sigma_1 \sigma_3$, the behaviors of coherence for the
$N$ qubits will be completely the same to that of the $N$ qubits
passing through the correlated phase flip channel.

(\romannumeral+3) For the correlated depolarizing channel, both
$\tilde{C}_{l_1}(\rho)$ and $\tilde{C}_{re}(\rho)$ can be obtained
directly by substituting the parameter $p$ in the expressions of
$\tilde{C}_{l_1}(\rho)$ and $\tilde{C}_{re}(\rho)$ for the phase
flip channel to $2p/3$, so their dynamical behaviors are
qualitatively the same to that of the phase flip channel.

\section{Total correlation gain and unlocalized quantum coherence} \label{sec:4}
In this section, we investigate the interplay between coherence of
the $N$-qubit system and its correlation created by the correlated
channel. It has been shown that unlocalized coherence, i.e., the
coherence of a multipartite system subtract the coherence localized
in each of its subsystem, captures a kind of correlation
\cite{Tan,RQC}. For the input state $|\Psi_N\rangle$, there is no
any correlation in it at the initial time. For $\mu=0$, from Eq.
\eqref{eq3a-4} one can obtain $C_{re}(\rho)=\sum_i C_{re}(\rho_i)$,
with $\rho_i$ being the reduced density operator of the $i$th qubit.
Thus there is no correlation to be created in the $N$-qubit system
when the phase flip channel is uncorrelated. But if $\mu>0$, the
consecutive applications of the channel turn to be classically
correlated and its correlation may be transferred to the $N$-qubit
system. For example, when $\mu=1$ the unlocalized quantum coherence
can be calculated as
\begin{equation}\label{eq4-1}
 C_{re}^{\mathrm{uqc}}(\rho)=C_{re}(\rho)-\sum_i C_{re}(\rho_i)= (N-1)H_2(p),
\end{equation}
and it is direct to show that $C_{re}^{\mathrm{uqc}}(\rho)$ equals
the quantum mutual information $I(\rho)=\sum_i S(\rho_i)-S(\rho)$,
which characterizes the total correlation contained in a system
\cite{rmp-qd}.

In effect, for the input state $|\Psi_N\rangle$ and the four
channels considered here, we always have $S(\Delta(\rho))=N$ and
$S(\Delta(\rho_i))=1$ ($\forall~i$), irrespective of the value of
$\mu$. So $C_{re}^{\mathrm{uqc}}(\rho)= I(\rho)$, namely,
$C_{re}^{\mathrm{uqc}}(\rho)$ equals the total correlation gain of
the $N$ qubits after they passing through the correlated channel.
One can further show that $C_{re}^{\mathrm{uqc}}(\rho)$ is an
increasing function of $\mu$. As $\mu$ quantifies the amount of
classical correlation in the implementation of the channel
\cite{cc1}, this result implies that the more classical correlation
there is in the correlated channel, the more correlation will be
transferred to the $N$-qubit system.

Moreover, if one concentrates only on the correlated phase flip
channel, then for any multiqubit input state $\rho_0$, its diagonal
part will remains unaffected. So one can obtain
\begin{equation}\label{eq4-2}
 C_{re}^{\mathrm{uqc}}(\rho)- C_{re}^{\mathrm{uqc}}(\rho_0)=I(\rho)-I(\rho_0).
\end{equation}
It indicates that the change of the unlocalized quantum coherence
for the multiqubit system equals its total correlation gain. To be
explicit, there are correlations created by the correlated channel
in the multiqubit system only when $I(\rho)>I(\rho_0)$.

\section{Conclusion} \label{sec:5}
In conclusion, we have explored decoherence process of $N$ qubits
when they pass through a correlated noisy channel. We used two
coherence measures, i.e., the normalized $l_1$ norm of coherence and
relative entropy of coherence, and considered four common noise
sources: phase flip, bit flip, bit-phase flip, and depolarizing
channels. Based on these, we analyzed in detail effects of the
consecutive applications of the noisy channel on decoherence of the
$N$-qubit system that is prepared initially in the maximally
coherent state.

We observed here two distinctive phenomena which may be exploited
for preserving coherence of a $N$-qubit system. First, both the
decay of the $l_1$ norm of coherence and the relative entropy of
coherence were suppressed if the consecutive actions of the channel
has some correlations. Second, the two coherence measures show
different dependence on qubit number $N$ of the system. The $l_1$
norm of coherence does not behave as a monotonic function of $N$,
and apart from the situation that the channel is fully correlated,
its value will becomes infinitesimal when $N\rightarrow \infty$.
However, the relative entropy of coherence becomes more and more
robust with the increase of $N$, and even when $N$ approaches
infinite, it still maintains a considerable value. This indicates
that the relative entropy of coherence for a system with large
number of qubits has an intrinsic rigidity against detrimental
effects of the channel.

To give an explanation for the delayed decoherence, we further
explored the interplay between the unlocalized quantum coherence and
the correlation gain of the $N$ qubits. The results show that for
the input maximally coherent state and the four sources of noise
considered here, the change of the unlocalized quantum coherence for
the $N$ qubits equals their total correlation gain. Moreover, for
the correlated phase flip channel, the change of the unlocalized
quantum coherence equals the total correlation gain for arbitrary
$N$-qubit input state. This implies that the delayed decay of
coherence may be caused by the inflow of correlations from the
channel to the system.

\section*{ACKNOWLEDGMENTS}
This work was supported by National Natural Science Foundation of
China (Grants No. 11675129, No. 91536108, and No. 11774406),
National Key R \& D Program of China (Grants No. 2016YFA0302104 and
No. 2016YFA0300600), the New Star Project of Science and Technology
of Shaanxi Province (Grant No. 2016KJXX-27), the Strategic Priority
Research Program of Chinese Academy of Sciences (Grant No.
XDB28000000), and the New Star Team of XUPT.

\newcommand{\PRL}{Phys. Rev. Lett. }
\newcommand{\RMP}{Rev. Mod. Phys. }
\newcommand{\PRA}{Phys. Rev. A }
\newcommand{\PRB}{Phys. Rev. B }
\newcommand{\PRE}{Phys. Rev. E }
\newcommand{\PRX}{Phys. Rev. X }
\newcommand{\NJP}{New J. Phys. }
\newcommand{\JPA}{J. Phys. A }
\newcommand{\JPB}{J. Phys. B }
\newcommand{\PLA}{Phys. Lett. A }
\newcommand{\NP}{Nat. Phys. }
\newcommand{\NC}{Nat. Commun. }
\newcommand{\SR}{Sci. Rep. }
\newcommand{\EPJD}{Eur. Phys. J. D }
\newcommand{\QIP}{Quantum Inf. Process. }
\newcommand{\QIC}{Quantum Inf. Comput. }
\newcommand{\AoP}{Ann. Phys. }
\newcommand{\PR}{Phys. Rep. }
%


\begin{thebibliography}{50}
\bibitem{Ficek} Z. Ficek and S. Swain, \textit{Quantum Interference and Coherence: Theory and Experiments}, Springer Series in Optical Sciences (Springer, Berlin, 2005).
\bibitem{Plenio} A. Streltsov, G. Adesso, and M. B. Plenio, \RMP {\bf 89}, 041003 (2017).
\bibitem{Hu} M. L. Hu, X. Hu, J. C. Wang, Y. Peng, Y. R. Zhang, and H. Fan, \PR {\bf 762-764}, 1 (2018).
\bibitem{Nielsen} M. A. Nielsen and I. L. Chuang, \textit{ Quantum Computation and Quantum Information} (Cambridge University Press, Cambridge, UK, 2000).
\bibitem{metro1}V. Giovannetti, S. Lloyd, and L. Maccone, Science {\bf 306}, 1330 (2004).
\bibitem{roc1} C. Napoli, T. R. Bromley, M. Cianciaruso, M. Piani, N. Johnston, and G. Adesso, \PRL {\bf 116}, 150502 (2016).
\bibitem{subc} K. Bu, U. Singh, S. M. Fei, A. K. Pati, and J. Wu, \PRL {\bf 119}, 150405 (2017).
\bibitem{biology} N. Lambert, Y.-N. Chen, Y.-C. Cheng, C.-M. Li, G.-Y. Chen, and F. Nori, \NP {\bf 9}, 10 (2013).
\bibitem{coher} T. Baumgratz, M. Cramer, and M. B. Plenio, \PRL {\bf 113}, 140401 (2014).
\bibitem{coher-ent} A. Streltsov, U. Singh, H. S. Dhar, M. N. Bera, and G. Adesso, \PRL {\bf 115}, 020403 (2015).

\bibitem{roc2} M. Piani, M. Cianciaruso, T. R. Bromley, C. Napoli, N. Johnston, and G. Adesso, \PRA {\bf 93}, 042107 (2016).
\bibitem{convex1} X. Yuan, H. Zhou, Z. Cao, and X. Ma, \PRA {\bf 92}, 022124 (2015).
\bibitem{convex2} A. Winter and D. Yang, \PRL {\bf 116}, 120404 (2016).
\bibitem{convex3} X. Qi, T. Gao, and F. Yan, \JPA {\bf 50}, 285301 (2017).
\bibitem{lqo1} X. Hu, A. Milne, B. Zhang, and H. Fan, \SR {\bf 6}, 19365 (2016).
\bibitem{lqo2} Y. Yao, G. H. Dong, L. Ge, M. Li, and C. P. Sun, \PRA {\bf 94}, 062339 (2016).
\bibitem{lqo3} M. L. Hu, S. Q. Shen, and H. Fan, \PRA {\bf 96}, 052309 (2017).
\bibitem{lqo4} C. L. Liu, Y. Q. Guo, and D. M. Tong, \PRA {\bf 96}, 062325 (2017).
\bibitem{lqo5} M. Takahashi and E. Chitambar, \JPA {\bf 51}, 414003 (2018).
\bibitem{power1} A. Mani and V. Karimipour, \PRA {\bf 92}, 032331 (2015).

\bibitem{power2} P. Zanardi, G. Styliaris, and L. Campos Venuti, \PRA {\bf 95}, 052306 (2017).
\bibitem{power3} K. Bu, A. Kumar, L. Zhang, and J.Wu, \PLA {\bf 381}, 1670 (2017).
\bibitem{power4} G. Styliaris, L. Campos Venuti, and P. Zanardi, \PRA {\bf 97}, 032304 (2018).
\bibitem{power5} H. Situ and X. Hu, \QIP {\bf 15}, 4649 (2016).
\bibitem{power6} L. Zhang, Z. H. Ma, Z. H. Chen, and M. Fei, \QIP {\bf 17}, 186 (2018).
\bibitem{break} K. Bu, Swati, U. Singh, and J. Wu, \PRA {\bf 94}, 052335 (2016).
\bibitem{ncgc} X. Hu, \PRA {94}, 012326 (2016).
\bibitem{steer} X. Hu and H. Fan, \SR {\bf 6}, 34380 (2016).
\bibitem{naqc1} D. Mondal, T. Pramanik, and A. K. Pati, \PRA {\bf 95}, 010301(R) (2017).
\bibitem{naqc2} M. L. Hu and H. Fan, \PRA {\bf 98}, 022312 (2018).

\bibitem{naqc3} M. L. Hu, X. M. Wang, and H. Fan, \PRA {\bf 98}, 032317 (2018).
\bibitem{fro1} T. R. Bromley, M. Cianciaruso, and G. Adesso, \PRL {\bf 114}, 210401 (2015).
\bibitem{fro2} X. D. Yu, D. J. Zhang, C. L. Liu, and D. M. Tong, \PRA {\bf 93}, 060303(R) (2016)
\bibitem{fro3} I. A. Silva \textit{et al.}, \PRL {\bf 117}, 160402 (2016).
\bibitem{fro4} A. Zhang, K. Zhang, L. Zhou, and W. Zhang, \PRL {\bf 121}, 073602 (2018).
\bibitem{fact} M. L. Hu and H. Fan, \SR {\bf 6}, 29260 (2016).
\bibitem{dyna1} Y. J. Zhang, W. Han, Y. J. Xia, Y. M. Yu, and H. Fan, \SR {\bf 5}, 13359 (2015).
\bibitem{dyna2} X. B. Liu, Z. H. Tian, J. C. Wang, and J. L. Jing, \AoP {\bf 366}, 102 (2016).
\bibitem{dyna3} W. Wu and J. Q. Cheng, \QIP {\bf 17}, 300 (2018).
\bibitem{dyna4} G. Guarnieri, M. Kol\'{a}\v{r}, and R. Filip, \PRL {\bf 121}, 070401 (2018).

\bibitem{dyna5} C. Mukhopadhyay, \PRA {\bf 98}, 012102 (2018).
\bibitem{cc0} F. Caruso, V. Giovannetti, C. Lupo, and S. Mancini, \RMP {\bf 86}, 1203 (2014).
\bibitem{cc1} C. Macchiavello and G. M. Palma, \PRA {\bf 65}, 050301(R) (2002).
\bibitem{cc2} C. Addis, G. Karpat, C. Macchiavello, and S. Maniscalco, \PRA {\bf 94}, 032121 (2016).
\bibitem{ordering} C. L. Liu, X. D. Yu, G. F. Xu, and D. M. Tong, \QIP {\bf 15}, 4189 (2016).
\bibitem{Tan} K. C. Tan, H. Kwon, C.-Y. Park, and H. Jeong, \PRA {\bf 94}, 022329 (2016).
\bibitem{RQC} M. L. Hu and H. Fan. \PRA {\bf 95}, 052106 (2017).
\bibitem{rmp-qd}K. Modi, A. Brodutch, H. Cable, T. Paterek, and V. Vedral, \RMP {\bf 84}, 1655 (2012).


\end{thebibliography}
%

\end{document}